# Strategies for spectroscopy on Extremely Large Telescopes: II – Diverse field spectroscopy

G.J. Murray and J.R. Allington-Smith[*]

Centre for Advanced Instrumentation, Physics Department,

Durham University, South Rd, Durham DH1 3LE

## Abstract

The fields of view of Extremely Large Telescopes will contain vast numbers of spatial sampling elements (spaxels) as their Adaptive Optics systems approach the diffraction limit over wide fields. Since this will exceed the detection capabilities of any realistic instrument, the field must be dilutely sampled to extract spectroscopic data from selected regions of interest. The scientific return will be maximised if the sampling pattern provides an adaptable combination of separated independent spaxels and larger contiguous sub-fields, seamlessly combining integral-field and multiple-object spectroscopy. We illustrate the utility of this Diverse Field Spectroscopy (DFS) to cosmological studies of galaxy assembly. We show how to implement DFS with an instrument concept: the Celestial Selector. This integrates highly-multiplexed monolithic fibre systems (MFS) and switching networks of the type currently available in the telecommunications industry. It avoids bulky moving parts, whose limitations were noted in Paper I. In Paper III we will investigate the optimisation of such systems by varying the input-output mapping.

# 1 Introduction

This is the second part of a study of strategies to produce instruments for Extremely Large Telescopes (ELTs) that are both affordable and efficient before the full benefits of Adaptive Optics delivering near-diffraction-limited images over a wide field become available.

The first paper (Allington-Smith, 2007; Paper I) discussed the potential for reducing instrument size through the use of image-slicing and the evolution towards integral field spectroscopy (IFS) as the AO system delivers improved image quality. This paper deals with the sampling strategies needed for key studies by ELTs and shows how this may be implemented using existing and emergent technologies.

The paper is laid out as follows. The case for Diverse Field Spectroscopy is made in §2 by reference to a particular example in cosmology. This includes the derivation of a basic set of requirements by reference to a meta-instrument, the Celestial Selector. In §3, we describe how fibre technology may be used to realise part of the system. In §4 we describe the use of optical switching networks similar to those currently available for the telecommunication industry. In Paper III, we discuss input-output mapping strategies which will enhance their performance.

---

[*] Email address: j.r.allington-smith@durham.ac.uk

# 2 Sampling the field of view

## 2.1 Dilute sampling

The problem of addressing the large number of data samples making up a datacube (voxels) obtained via a "3D" technique such as IFS (e.g. Allington-Smith 2005) with a finite number of detector pixels is well known[†]. At the diffraction limit, the number of potential spaxels available at the focal surface is ~$10^{10}$, requiring ~$10^{13}$ voxels, each of which should be addressed by at least $2^3$ detector pixels for optimum sampling at the Nyquist-Shannon limit. This implies an unaffordable detector array of $10^7$ x $10^7$ pixels. From this it follows that dilute sampling by a factor $10^{-6} - 10^{-4}$ of the focal plane is required.. This problem is accentuated in the near-infrared where detector costs are higher.

This can be achieved by directing light from movable sub-fields into a number of spectrographs which may be independently optimised as required. An example is the pickoff scheme used in KMOS (Sharples et al. 2005) in which 24 sub-fields of a few arcsec are directed by articulated arms containing an optical relay to fixed image slicing integral field units (IFUs) and from there into three spectrographs. Each sub-field covers roughly 400 spaxels and 400,000 voxels, or $10^4$ spaxels and $10^7$ voxels in total equivalent to a dilution factor of ~$3x10^{-6}$.

For an ELT, the same approach is beset with difficulties because of the greatly increased requirement on arm stiffness (see Paper 1). Even if mounted on a gravitationally-invariant platform, differential flexure between arms has the potential to cause errors when the field configuration is changed. Alternative solutions involve robot-positioned pickoff mirrors feeding optics arranged at the periphery of the field but these require complex optical trains that may also experience flexure problems. Apart from the great technical challenge, such schemes lack versatility since the IFU size is either fixed or crudely quantised. For example, this would make integrated spectroscopy of multiple faint objects inefficient unless the surface density was very high since each target would be addressed by a complete IFU. A further restriction is imposed by the minimum inter-field distance governed by the finite size of the pickoff mechanism, the limitations of the control system and arm geometry and the likelihood of unwanted field rotations as the pickoff position changes.

## 2.2 Diverse-field spectroscopy

Although the case for dilute sampling of the field is clear, the best strategy for sampling regions of interest within the field is not obvious. Another major uncertainty is the optimum spaxel size. Both issues are entirely dependent on the nature of the investigation. For example, measurement of intermediate-mass black holes in globular clusters requires a different spaxel size to that of high-redshift dark-energy surveys. The same is true of how those spaxels are distributed over the field. The range of possible investigations is so large that only very general statements can be made.

---

[†] A spaxel is the element of *sampling* in two spatial dimensions and a voxel is the sampling element in three dimensions (generally two spatial dimensions plus wavelength). These are related to the corresponding r*esolution* in these quantities by the *oversamping* which normally satisfies the Nyquist-Shannon sampling criterion that the FWHM of the distribution should be sampled by at least two detector pixels in each dimension. Thus the minimum required oversampling is 4 for a spaxel and 8 for a voxel.

(a) The *angular* spaxel size is likely to scale inversely with the telescope aperture (or, more accurately focal length) since this produces constant signal for extended sources and, to first-order, constant signal-to-noise ratio. The same trend applies for studies which exploit the improved diffraction limit of larger telescopes (for e.g. exo-planet studies) to scales of ~5mas. This means that the *physical* size of spaxels is likely to remain constant, independent of telescope aperture as already noted. Thus cosmological investigations with spaxel sizes currently 100-500mas might scale to 20-100mas on an ELT. A counteracting tendency arises from cosmological dimming of surface brightness which requires spaxel size $\propto (1+z)^{-2}$. Indeed detailed simulations indicate optimum spaxel sizes of 25-75 mas for this type of study (e.g. Evans et al. 2006). For the very faintest galaxies detected only in integrated light the optimum spaxel size is 100-300mas.

(b) The range of investigations to be carried out over a wide-field (from galactic archaeology to dark energy surveys) and the need to share wide-field exposures between different projects to maximise observing efficiency requires great flexibility in sampling patterns. This will range from single spaxels addressing individual objects so faint that only an integrated spectrum can be recovered, to large sub-fields in which the spaxels are both numerous and contiguous.

A completely flexible sampling pattern emerges as the most effective solution. This ranges from large contiguous sub-fields (perhaps a single field at the field centre) to many tiny independently-deployed sub-fields covering the whole telescope field. This combination of integral-field and multiple-object spectroscopy is known as diverse-field spectroscopy (DFS) and is characterised by its ability to generate completely arbitrary sampling patterns.

For the large contiguous fields the number of spaxels per field might be $10^4 < n < 10^6$, but $1 < n < 10$ for the greater numbers of very small fields,. There are also intermediate scales of interest such as that required to obtain virial galaxy masses at intermediate or large distances ($10^{2} < n < 10^3$) and survey regions for chemical studies in nearby galaxies and multiple resolved star-cluster studies ($10^3 < n < 10^4$ spaxels).

The pattern is subject to the limitation that

$$N_S = \sum n_i m_i = \frac{n_D Q}{n_\lambda}$$

where $n_i$ is the number of spaxels per sub-field of type $I$ of which there are $m_i$ of that type. $N_S$ is the number of spaxels available in total which is determined by the number of detector pixels $n_D$, and $n_\lambda$ is the number of pixels covering the spectrum simultaneously. $Q$ is the specific information density (Allington-Smith 2006) which depends on the technique used and the required independence of the spectra. For pure IFS, the spectra are allowed to overlap in the spatial direction since the Shannon-Nyquist sampling criterion applies at the IFU input, not at the slit (Allington-Smith & Content 1999) and it is possible to arrange for spaxels which are adjacent on the sky to be adjacent at the slit. However DFS violates this condition since adjacent spaxels may come from widely separated parts of the sky. In this case, $Q < 0.2$ to minimise cross talk.

Where discrete deployable pickoffs are used (arms, buttons, robot-bugs), there may also be a limit on the number of sub-fields, $N_f = \Sigma m_i$, imposed by the distance of closest approach packing densities.

## 2.3   Celestial Selector

The basic principle of an instrument for DFS is shown in Figure 1 and Figure 2. The celestial selector simply routes input light from the telescope focal surface to one of a number of outputs which feed spectrographs optimised for different types of spectroscopy. The concept consists of an input fibre bundle with a very large number of short, broadband fibres in a single monolithic unit, an array of optical switches which either permits light from selected inputs to go through to the output or redirects it into an available output channel; and a set of output bundles containing a much smaller number of fibres feeding a variety of spectrographs.

Although envisaged here as a fibre system it is possible to consider an alternative realisation in which either the input or output fibre systems are replaced by direct optical feeds. An earlier example of such a system is the *Honeycomb* system of Bland-Hawthorn et al (2004; discussed further below) in which output fibre bundles are plugged into a very large lenslet array. See also Bland-Hawthorn (2007) for a description of the Maximus concept.

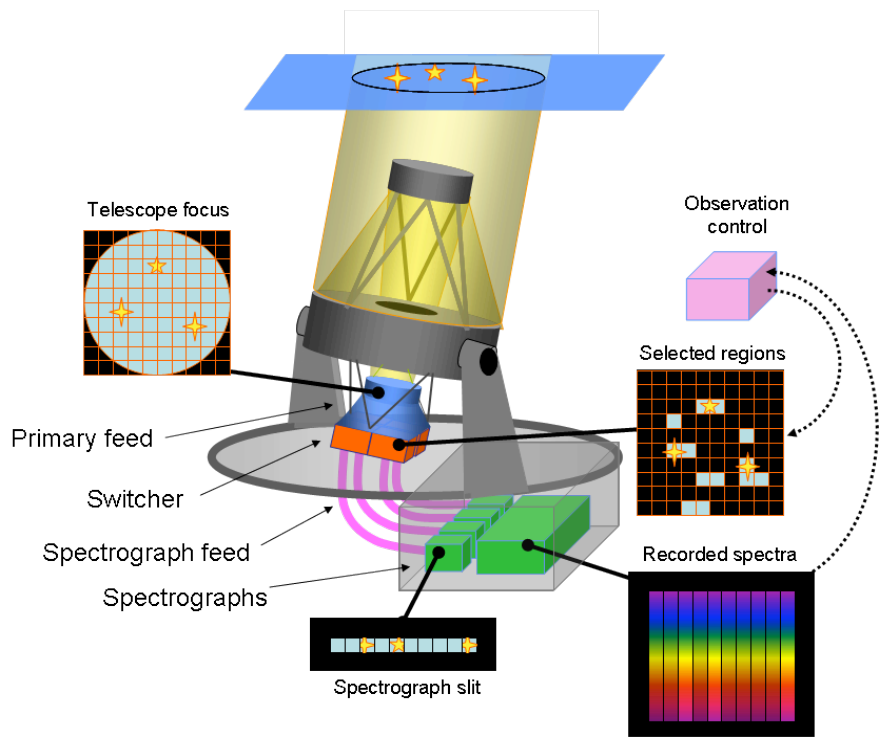

**Figure 1: The Celestial Selector concept.**

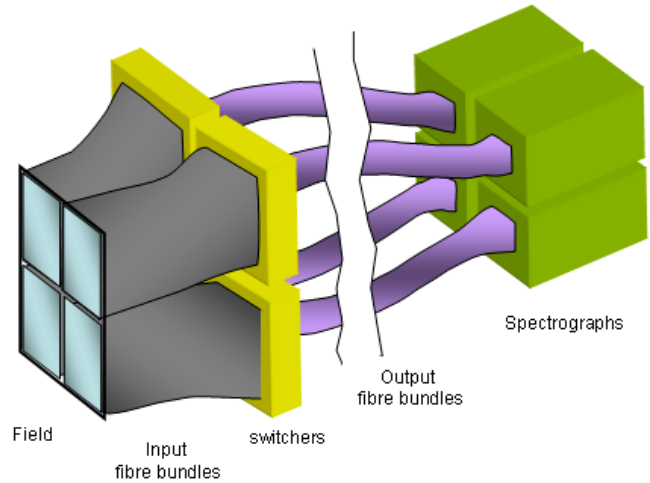

**Figure 2: Schematic of the *Celestial Selector* using fibre technology. This example has 4 tiles, each of which maps through a separate switcher (envisaged as a 2-way multiplexer) to a single spectrograph. See the text for details.**

## 2.4    A cosmological example

As an illustration, we consider an example of a cosmological study with an ELT. Figure 3 shows LAB-1, the field of Lyα-emitting objects at redshift z ~ 3 obtained with the SAURON IFU (Bacon et al. 2001) on the 4.2m Herschel telescope, collapsed in wavelength to form an image at the rest-frame Lyα wavelength (Weijmans et al. 2009; see also Bower et al. 2004). The spectral data reveals the kinematics of the clouds and details of the emission line shape to distinguish between different models. The luminosity of photoionisation and re-radiated mid-infrared emission places limits on the size and mass of these pre-galactic structures and the halo in which they are presumably embedded.

Although uncommon at the limits accessible to SAURON, it is possible that structures such as these will be representative of the universe at the flux levels accessible to an ELT. It is therefore illustrative to use this observation as a guide to the formats of interest to cosmological studies with ELTs.

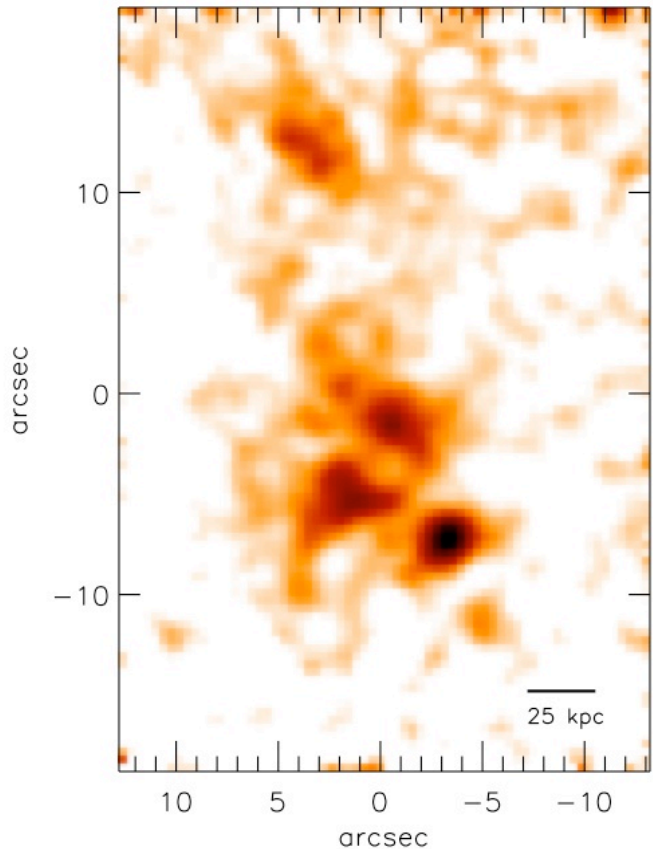

**Figure 3: Image of Lyα-emitting region LAB1 in the SSA 22 protocluster region at z = 3.09 (from Weijmans et al. 2009). The spatial extent is ~100kpc and the Lyα luminosity ~$10^{37}$W.**

The image covers 0.6 arcmin$^2$ and contains several objects embedded in an extensive nebulosity. The central complex of features has AB surface brightness ~27.5 mag/arcsec$^2$ and covers 200 arcsec$^2$. This implies a Lyα luminosity of ~$10^{37}$W over ~$10^4$kpc$^2$. Dividing the area of the complex by the number of blobs suggests that we need to assign a contiguous sub-field of ~6 arcsec to each resolved object. This is consistent with the isophotal areas of resolved Lyα clouds reported by (Matsuda et al. 2006) as 20 - 200 arcsec$^2$.

Figure 4 shows the results of scaling to an ELT with aperture 40m as a function of the spaxel size, reaching a limit of ~ 30 mag/arcsec$^2$. To be conservative, we have assumed that half of the objects are unresolved. The total numbers of resolved and unresolved objects accessible to an ELT have been predicted on the assumption of constant co-moving density with a range of standard cosmology ($\Omega_M + \Omega_\Lambda = 1$ with $0.3 < \Omega_M < 0.7$) for $3 < z < 8$, taking account of the different detectability of resolved and unresolved objects as a function of spaxel size.

The figure shows the number of active spaxels required to cover the regions of interest (RoIs) after scaling to the ELT sensitivity limit, and the fraction of the field covered by "active" spaxels. Also shown is the number of spaxels required to cover a complete contiguous field of $1 \times 1$ and $3 \times 3$ arcmin$^2$ for the surface density extrapolated from that of the SAURON data and for a more-representative global value, 10 times lower.

The global value was estimated from the data of Matsuda et al. who find similar objects to have a surface density of ~ 0.4 arcmin$^{-2}$ over much wider fields than the SAURON image.

Since the central complex in the image has Lyα flux $\sim 10^{-21}$ Wm$^{-2}$, similar to that of Matsuda et al. we conclude that the image of Weijman et al contains the equivalent of one Lyα emitter of the type surveyed by Matsuda et al. based on its flux, or several based on its spatial extension. Taking the equivalent source count in the SAURON field as 2 implies an overdensity with respect to the global value of $\sim 10$. This is broadly consistent with Le Delliou et al. (2006) who predict surface densities in the range 0.1-3 arcmin$^{-2}$ at $z \sim 8$ depending on reionisation epoch and escape fraction, for a redshift interval of 0.1. This compares with our extrapolation of $\sim 20$ arcmin$^{-2}$ at the ELT limit which is consistent with an overdensity of at least 10.

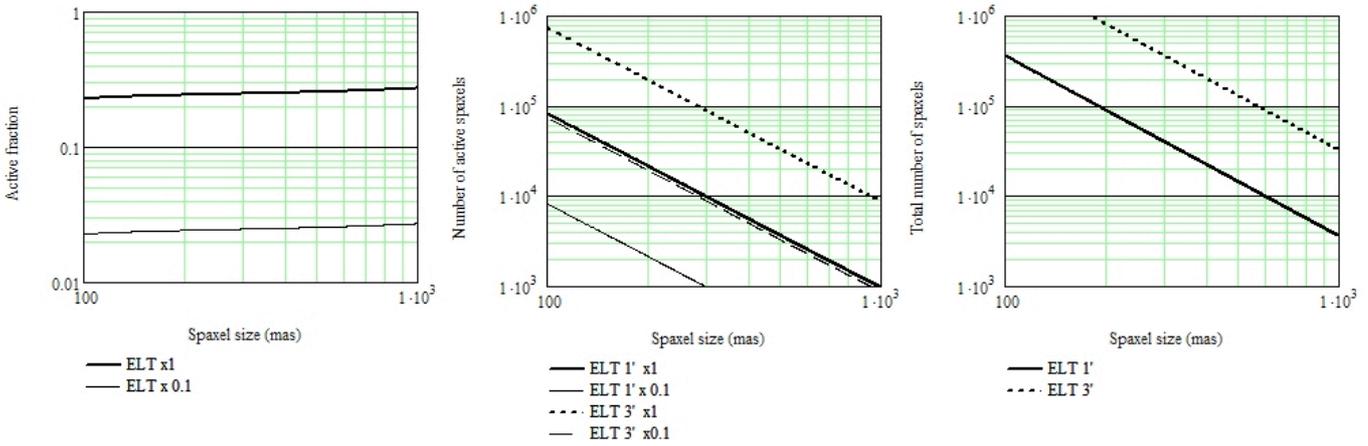

**Figure 4: Field format for a 40m telescope extrapolated from the example observation discussed in the text: Left - the fraction of the field covered by active spaxels; centre - the number of active spaxels; right - the number of spaxels required to cover the full field. The solid curves are for a field area of 1 arcmin and the dotted curves for 3 arcmin. Heavy lines are for the surface density implied by the example observation and normal weight indicates a surface density 10 times lower.**

**Table 1: Minimum spaxel sizes to cover all regions of interest in the field for the stated number of spaxels either active (input to the spectrograph) or total (to cover the full field).**

| Number of spaxels | $10^4$ | | $10^5$ | $3\times10^4$ | | $3\times10^5$ |
|---|---|---|---|---|---|---|
| Type of spaxel | Active | | Total | Active | | Total |
| Surface density normalisation | x0.1 | x1.0 | | x0.1 | x1.0 | |
| Field size = 1 arcmin | 100 mas | 300 mas | 200 mas | <100 mas | 180 mas | 110 mas |
| Field size = 3 arcmin | 300 mas | 1000 mas | 700 mas | **180 mas** | 500 mas | **350 mas** |

The figure shows that the fraction of the field covered is 2.5 - 25% for the smallest spaxel sizes. Table 1 shows what can be achieved for a specified number of active and total spaxels by stating the minimum spaxel size that will provide the required coverage. For example, consider the highlighted values of 180 mas and 350 mas: This indicates that 3 x 10$^4$ active will cover the defined RoIs provided that the spaxel size is >180 mas for a field size of 3 arcmin for the lower surface density normalisation and that this field can be covered by a total (active or otherwise) of 3 x $10^5$ spaxels, provided they are of size 350 mas or bigger.

It is instructive to consider if fields of this type can be satisfactorily addressed by either a single large IFU or by deployable mini-IFUs . For the single IFU we note that the fraction of the field containing active spaxel never exceeds 30% and ranges down to just a few percent implying inefficient use of the IFU field which, for the same number of active spaxels (i.e. those routed to the spectrograph) will be smaller by a factor 3-40. For deployable IFUs, if the same number of active spaxels is distributed between 50 mini-IFUs, this implies 200 spaxels per IFU. To match our requirement of 40 arcsec$^2$/blob implies spaxel sizes of at least 450mas. This is uncomfortably coarse but a bigger problem is that the fraction of the field covered by the pickoffs would exceed 50% for a 1-arcmin field at which level sub-field crowding is very likely to be a problem. The question of efficient mapping strategies is discussed further in Paper III where we simulate how this field could be addressed by the type of DFS technology discussed in the next section.

Thus, we conclude that this type of cosmological study would be well-served by a diverse-field spectrograph with ~$10^4$ spectral channels within a total field covered by ~$10^5$ spaxels, although increasing this format by a factor of a few would be beneficial. We shall take this as a working specification. Alternative solutions involving a single panoramic IFU or deployable mini-IFUs do not provide a good match to the distribution of regions of interest.

# 3 Technologies for DFS

## 3.1 Three stages

In discussing appropriate technologies, we must break the problem into three stages. Firstly the transport of light from the telescope focal surface; secondly the down-selection process; and thirdly the transport of light from there to the spectrographs. The third of these presents no unusual problems except for the need to minimise fibre length to preserve short-wavelength transmission. The only aspect that warrants discussion in this paper relates specifically to one of the proposed down-selection methods, and a scheme for ameliorating an inherently inefficient coupling between the downselector and the spectrograph fibre train.

We make the case (below) for using optical fibres for the first stage and discuss an implementation named MAIFU (Massively Addressable Integral field Unit). The Honeycomb concept of Bland-Hawthorn et al (2004) is briefly described as an illustration that fibres are not the only way to do this (although this uses fibres to feed the spectrographs). However, the bulk of the discussion concerns optical switching networks (§4).

## 3.2 Why use fibres?

Although other solutions are possible, it is clearly advantageous to minimise the number and size of any moving parts or at least to remove the requirement for high stability during an observation. Fibres have long been used in astronomy. The multimode fibres required have the disadvantage of being inherently lossy in terms of entropy; Etendue is not conserved, resulting in Focal Ratio Degradation (FRD) which either reduces performance (efficiency and/or resolution) for a fixed instrument cost or makes the instrument bigger and more expensive. Furthermore, their format limitations lead to a relatively low specific information density (Allington-Smith 2006) compared with slicing systems when used in IFS. However, they offer greater versatility than other types of optical relay, due to their flexibility and the ability to be arbitrarily rerouted. This attribute may be exploited to produce simpler instruments than those requiring articulated optical trains and is further desirable for instruments mounted off the telescope, to isolate them from the movement of the telescope during tracking. As we have seen (Paper 1), these issues are more important as the telescope aperture increases and so must be addressed for ELTs.

Developments in fibre technology also may make fibre systems more attractive. For example, photonic crystal fibres with large mode area (LMAs) may be used for transporting light coherently without modal diffusion or modal stripping - the sources of FRD - while providing satisfactory input coupling efficiency and a very broad effective bandwidth due to their *endless single mode* behaviour (Corbett et al. 2006).

Finally, the maturity of fibre systems for IFS allows us to contemplate the construction of very large fibre bundles which, as we will show, forms the hardware basis of the novel fibre systems presented below.

### 3.3 Monolithic fibre systems

The construction of very large fibre systems is illustrated by the MAIFU testbench proposed by CfAI. The system contains a coherent 2-D fibre array with high filling factor (using lenslet arrays) which defines the contiguous input field (Figure 5).

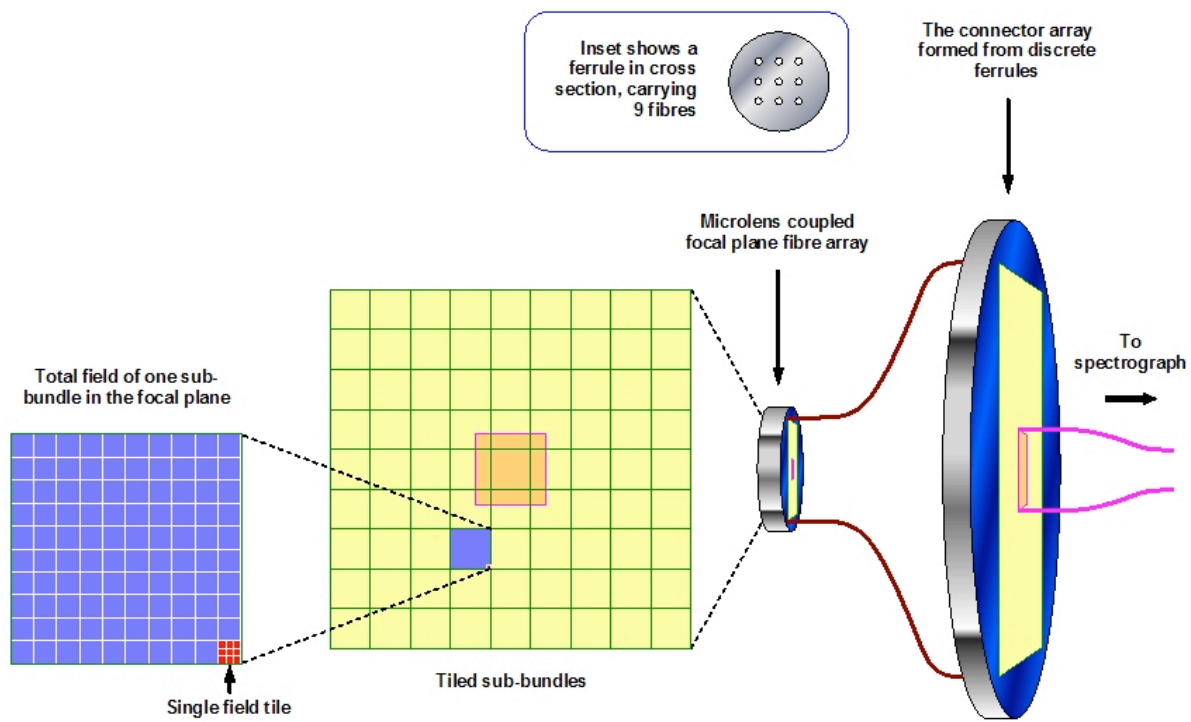

**Figure 5: The MAIFU concept.**

The fibres terminate in an array of connectorised breaks, the other side of which feeds through to the spectrographs. The downselection is effected by making fibre connections to the spectrographs via appropriately-sized bundles with each bundle positioned as required. Thus the end-to-end fibre system is re-configurable, subject to the limitations imposed by the output bundle size, shape and modularity.

The connector array is designed as a regular grid of miniature sub-connectors. Fibres are grouped into sets within each sub-connector by utilising discrete multi-fibre ferrules. This keeps the number of connectors to a reasonable level and defines subsets of fibres that map to corresponding minimum-sized *tiles* of the input field. Connectors thus form the smallest discrete sub-fields corresponding to an individual RoI. The field is thus effectively addressed by a large number of tiles, acting as small IFUs. A fibre-fed spectrograph can be plugged into regions of the field with the flexibility required for DFS, generating contiguous sub-fields ranging in the extremes from small and numerous to large and unique. The spatial information in the smallest fields reduces the need for accurate target acquisition, permitting a fixed array of sampling points to function in a high resolution multi-object mode.

The most conservative option for the reconfiguration is to use a conventional pick-and-place industrial robot. In this case, the pitch of the ferrules in the connector array has to be sufficient to allow access for the positioner. The number of fibres/ferrule is subject to an optimisation determined by the achievable fibre packing density and the maximum number of connectors that can be addressed in a reasonable time. The precision-ferrule connectors would be passively self-aligning in both translation and rotation, so the robotic mechanism needs only modest positioning precision. The entire instrument would be engineered in modular fashion from a number of tiled sub-bundles rather than one single monolithic device; modularity simplifies construction and permits a greater degree of design flexibility.

The Honeycomb concept (§2.3) is similar in principle but dispenses with the input fibres, replacing them with an array of lenslets that deliver a micropupil to the core of the output fibres. Although Honeycomb is optically-simpler, the MAIFU system is more flexible in that the pitch of the connector interface can be selected independently of the lenslet array and the interface can, in principle, be remote from the telescope (however in practical terms, a very long fibre bundle is undesirable). The flexibility in input/output pitch and the ability to change the input/output mapping is a further advantage of the MAIFU concept.

The limiting factor for both systems is the downselection method. Using a robot, the reconfiguration time could be a problem if the number of connectors (i.e. sub-fields) is large, although the job could be shared by multiple robots. A related problem is that the connector array could become unfeasibly large as the fibre count is increased. For these reasons, as well as the general desire to minimise or miniaturise moving parts, we now consider the use of arrays of optical switches to perform the downselection.

# 4 Switching systems for DFS

## 4.1 Switching geometry definition

We may define the geometry of a MAIFU-type fibre system of the type shown in Figure 2 as follows: There are $N_T$ tiles each containing $N_C$ cells of $N_S$ spaxels. Within each cell, $N_D$ spaxels can be routed to the output. Therefore the total number of inputs and outputs and the downselection factor respectively are defined by

$$N = N_T N_C N_S \quad M = N_T N_C N_D \quad G \equiv \frac{N}{M} = \frac{N_S}{N_D} \tag{1}$$

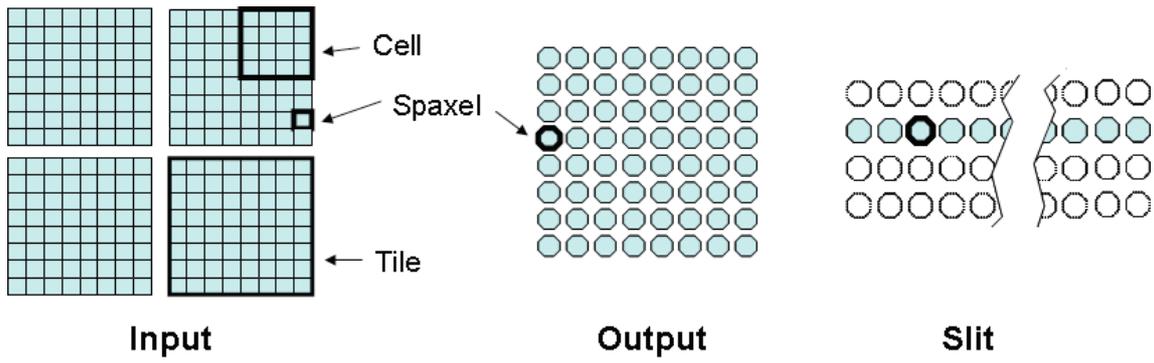

**Figure 6: Fibre system geometry. In this example there are 4 tiles each containing 4 cells with 16 spaxels in each. Of these, 4 can be routed to the output so there are 64 possible outputs. The output format can be independent of the input in geometry, scale and mapping. The layout at the slit includes the option of slit multiplexing (by a factor of 4) as discussed in the text. Dispersion is in the vertical direction.**

The example in Figure 2 has $N_T = 4$ tiles, $N_C = 4$ switcher/tile, $N_S = 16$ spaxels/cell and $N_D = 1$ outputs/cell so that $N = 256$, $M = 16$ and $G = 16$.

## 4.2 One-way multiplexing with spatial light modulator arrays

An example of a device which will route one of many inputs to a single output is a spatial light modulator (SLM). This can block light from all but one input so $N_D = 1$.

Inevitably such devices are unable to address contiguous RoIs since fibres which are adjacent in the field will be routed to the same cell from which only one can be routed onward to the slit. Only if the RoI is at the edge of two cells or at the corner of 4 can this problem be overcome. For this reason, an incoherent mapping scheme is adopted such that there is a finite probability that two spaxels adjacent in the field will be directed to different cells with the possibility that both may be routed to the slit. The example system shown in Figure 7 has this feature built in. The optimum remapping strategy is investigated in Paper III.

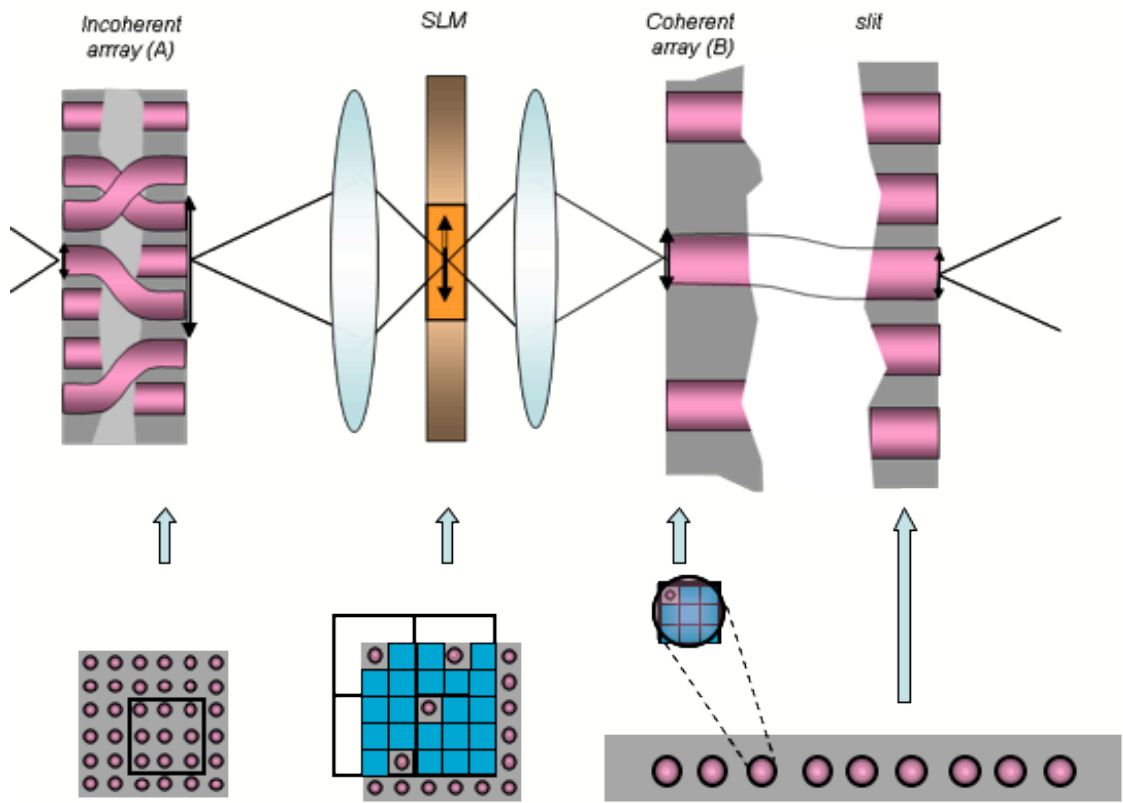

**Figure 7: Illustration of the principle of an incoherently-reformatting integral field unit using a one-way multiplex switcher (a spatial light modulator). The field of view is reimaged on the face of a fibre array which is then remapped according to a predefined scheme. The remapped fibres are input in small groups to an array of cells in a spatial light modulator which passes light from one user-selected fibre in each cell. The output is fed into a reduced number of fibres which are reformatted (without randomisation) into the one-dimensional slit as in a conventional integral field unit.**

In this example, the modulator is sandwiched between the input and output fibre bundles which have different pitch and fibre core size. This relies on the scrambling of the fibre to homogenise the input light so that approximately the same amount of light is transmitted to the output regardless of which element of the cell is open. This implies a non-conservation of Etendue resulting in loss of performance/cost and is therefore sub-optimal, but may be cost-effective if the technology is readily available. The modulator array itself could use various technologies including MEMS microshutter, pixellated liquid crystals or lithium niobate.

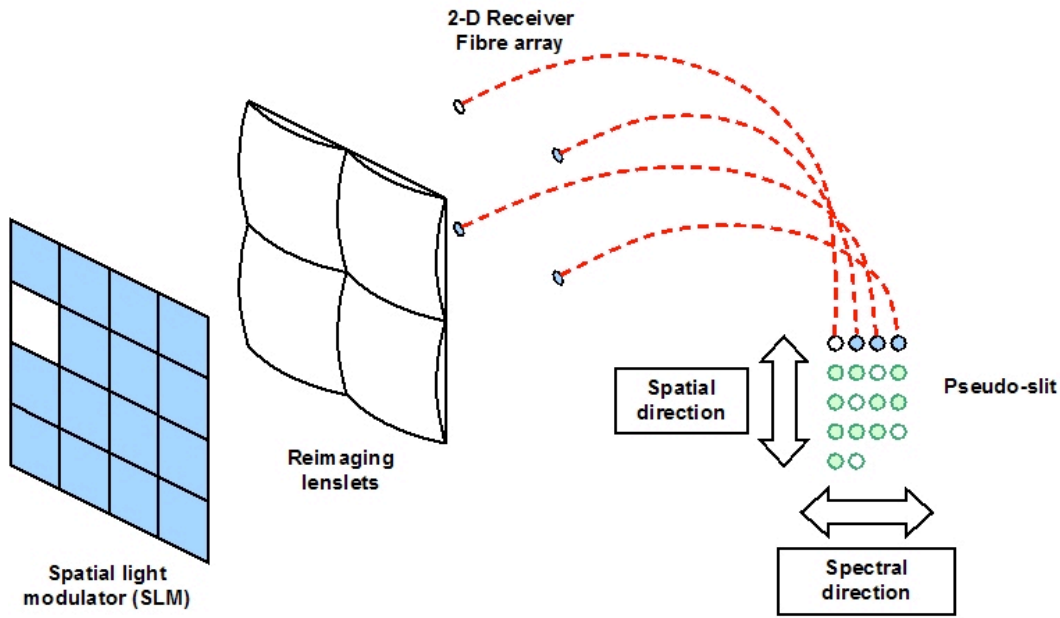

**Figure 8: Sketch of SLM cell modified to optimise performance.**

To reduce issues of homogeneity and FRD, the system may be modified by providing additional fibres to address each cell. In this example (Figure 8) a cell with $N_S = 16$ inputs is modified to provide $N_D = 4$ potential outputs; but only one of these is used. The selected spaxel produces a spectrum that can be displaced in the dispersion direction. This reduces off-axis angles which are now more consistent (and scrambled in azimuth) resulting in reduced FRD and light loss. The spectra from adjacent cells are now displaced from each other in wavelength so it is necessary to ensure that there is minimum cross-talk between adjacent fibres at the slit.

The advantage is that the required technologies already exist, and there are no moving parts. The disadvantage is that remapping is required, which complicates the construction of the fibre bundle.

### 4.3 Architectures for two-way multiplexing

Optical switching technology has advanced rapidly in recent years, to meet the growing demand for such devices in the telecommunications sector. Many novel concepts are in development or have been recently launched onto the commercial market. Three, very distinct approaches are emerging (see also Truex et al. 2003).

(a) **Beam-steering using optically-active materials.** They are dependent on the bulk optical properties of the active materials (e.g. Lithium Niobate) with a resultant compromise in throughput over a sufficiently broad band. They are generally tailored to applications requiring high switching speeds (~1 ns) and narrow spectral bandwidth (e.g. 10 nm at 1.55 μm with an insertion loss of 5 dBs, for a JDSU 2 × 2 channel commercial product). Therefore they are not optimal for DFS, although as an interesting aside, a comparatively efficient (70% throughput) 1×35 channel single mode system has been developed recently for NASA by ADVR Inc for a LIDAR application[‡]

---

[‡] See: http://www.advr-inc.com/switch.html for a description.

(b) **Fibre manipulation.** This varies both the angle and position of fibres to switch light paths. These devices differ widely in design and many systems display a high degree of ingenuity. They are also potentially highly efficient. A good example is the switching technology developed by Polatis Inc. where lensed fibres are steered directly by means of piezoelectrically driven fingers via a central optical relay. The single mode commercial products available from this company have switching speeds of <10 ms and the bandwidth is limited only by the modal restriction of the fibre, so they can function throughout the optical spectrum. The number of optical interfaces is small, (six including the fibre-lens interfaces) so the insertion losses are low, ~1 dB. There is to-date no plan to commercialise a multi-mode system however. Interesting as such schemes are, they are more suited to low fibre counts because the positioning actuators tend to be bulky and elaborate, incorporating relatively large mechanisms.. Again therefore they are not currently an ideal DFS solution.

(c) **Beam-steering using MEMS micromirror arrays**. These components are versatile, potentially of flexible format, and they can attain high packing densities so are suited to large fibre counts. The arrays are also reasonably inexpensive to produce and the manufacturing process (photolithographic replication) is such that low volume production of custom devices does not dramatically inflate the unit-cost. A switching scheme usually consists of a dual-array configuration with an intermediate (reflective) optical relay, in order that the displacement and angle of the relayed beams can be adjusted. Again therefore the systems are potentially efficient since the traversing light typically intercepts three reflective and four refractive surfaces only (including fibre-lens interfaces). The wavelength range is limited purely by the fibre and the design of the relaying microlenses that couple light into and out of the system. Commercial single mode switches are available, and multimode systems are in development by CrossFiber Inc and CfAI.

For our specific application, it seems clear that MEMS based micromirror arrays presently offer the best prospects for DFS.

## 4.4 Two-way multiplexing with micromirror arrays

Binary micro-mirror (BMM) arrays such as the spatial light modulators employed in digital projector systems have found use in astronomical spectroscopy. They either direct light into the spectrograph or direct it elsewhere, by switching between two fixed angles. The individual mirror size is ~0.1mm with a high filling factor obtained by close packing. However the tilt-precision achievable by individual micromirrors is relatively low and surface errors may result in scattered light and aberrations. In certain configurations they can act as blazed surface-relief gratings resulting in further stray light.

The other type of MEMS micromirror array employs fully-steerable micro-mirrors (SMM) with high angular precision (< 0.1mrad over ± 150mrad). It is understood that these components have as yet not seen any application in astronomical instrumentation (although the technology is relatively new). In such devices the mirrors are larger (0.3-1 mm typically) and the pitch is larger than the mirror size due to the mirror mounting flexures and electrical connections (Figure 9).

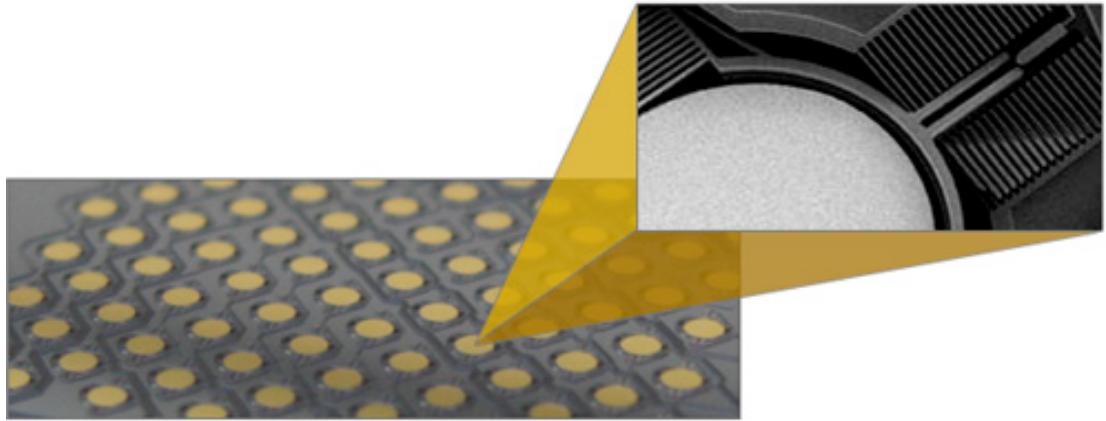

**Figure 9: A fully steerable MEMS based micromirror array (courtesy of CrossFiber Inc.).**

Figure 10 illustrates how this technology might apply to DFS. SMM arrays are used to select regions from the focal plane. Although the arrays are typically designed to work with fibre-optic input and output channels, the simplest, ideal Celestial Selector scheme would avoid a fibre-optic front-end entirely, and the spaxellated focal plane would be relayed directly onto a close-packed arrangement of micromirror arrays.

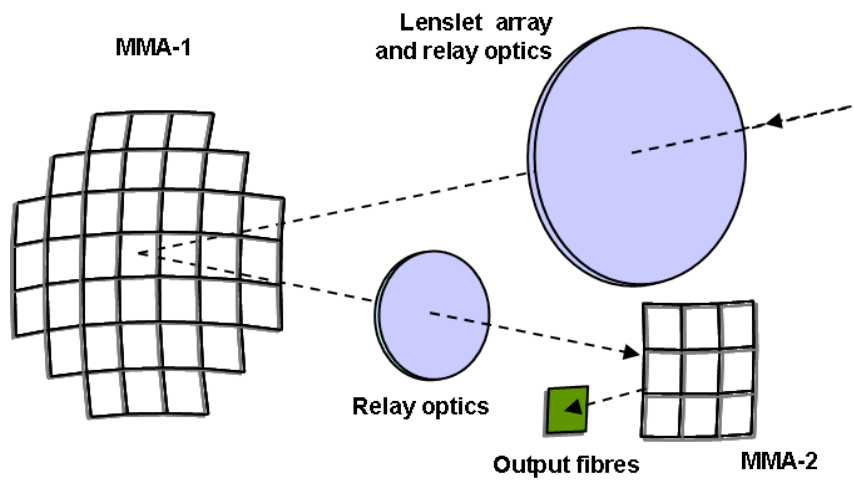

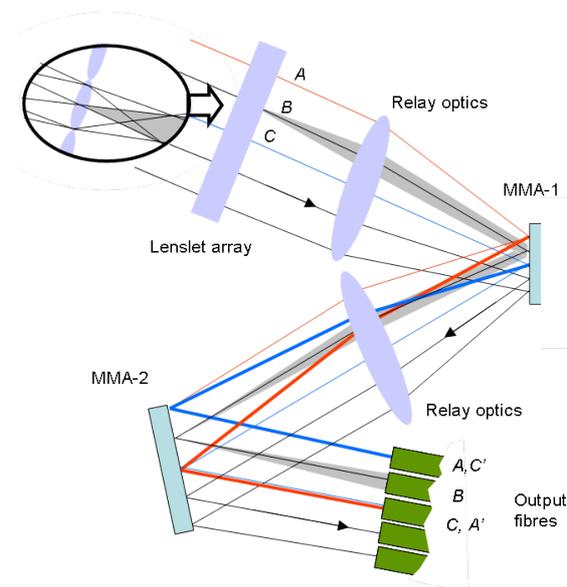

**Figure 10. Possible use of micromirror arrays (MMAs) in the Celestial selector. Top - general layout: Light enters from top-right and is re-imaged onto the primary micromirror array. The redirected light is then reimaged onto a location within the output fibre bundle by the second micromirror array. Bottom – optical principle showing principal rays and the envelope of rays incident at the centre of the lenslet for ray path B only. Rays A and C are shown with straight-through routing and with their positions interchanged (heavy lines); A', C') by appropriate tilting of each pair of micromirrors.**

The primary reformatting surface would consist of an array of micromirrors which select arbitrarily regions within the entire field. It can be seen that such an optical system will always require two mirror arrays to select the required position in the field and to correct the angle of the rays to avoid vignetting by the spectrograph stop.

The primary selection may be achieved more crudely by utilising a BMM array instead. In which case, the primary selector surface would now be sited directly at the focal plane (no lenslets). However, as discussed earlier, due to the optical limitations of such arrays the performance could be significantly compromised.

## 4.5   Maximising field size

The Celestial Selector concept is undoubtedly well-optimised for DFS but there are some potential limitations

- As already noted, the SMMs are ~1mm in size on a pitch of 2-3mm, and so are significantly larger than the BMM Thus a DFS system as envisaged above will be impractically large. However, there is no fundamental reason why the SMM array cannot be miniaturised – there simply has been no commercial incentive to do this.

- For optimum optical performance the optical path lengths need to be quite large (compared to the mirror apertures) which results in a large device and impractical accuracy in establishing and maintaining tilt angles.

- Fibre-fed switch modules as commercially supplied are designed as closed units, able to accommodate a finite and relatively low fibre count ($n \ll 10^3$). This by necessity will result in a design that divides the focal plane into a number of sub-fields, within which a fixed fraction of the spaxels can be coalesced into a contiguous region. This effectively limits the size of the largest possible contiguous RoI to the dimensions of the sub-fields.

To increase the field size and, specifically, the largest possible contiguous RoI, it is possible to introduce another switching layer, as illustrated in Figure 11, to allow the individual contiguous RoIs in each subfield to be merged into a larger one at the final output. This versatility would be bought at the price of reduced throughput but might nevertheless maximise the figure of merit for the investigation in question.

The use of MEMS optical switching for downselection confers two additional capabilities which make it a unique solution for DFS:

(a) **Variable resolution.** Light from adjacent spaxels in a RoI may be combined and fed into a single receiver fibre, effectively making a 'super-spaxel' with larger angular size and correspondingly greater light gathering power (Figure 12). Thus regions may be switched between small-field, high resolution, and large-field, low resolution with boosted signal sensitivity. However this makes two assumptions: firstly, that the array has not been remapped; secondly, that the implied non-conservation of Etendue matters less than the benefit of increased sampling versatility. Clearly when selecting the fibre numerical aperture, the system must be designed with this capability in mind, in order to minimise potential losses due to non-optimal beam propagation resulting from the faster incident beam.

(b) **Variable, independent exposure times.** During an exposure of a fixed field, the objects within that field are not restricted to a common exposure time. The optical interconnection can be broken at any time to terminate an exposure, or it can be reconfigured to target another object. This is impossible with robotic systems because they interrupt the light path of other fibres during reconfiguration and mechanically perturb the instrument

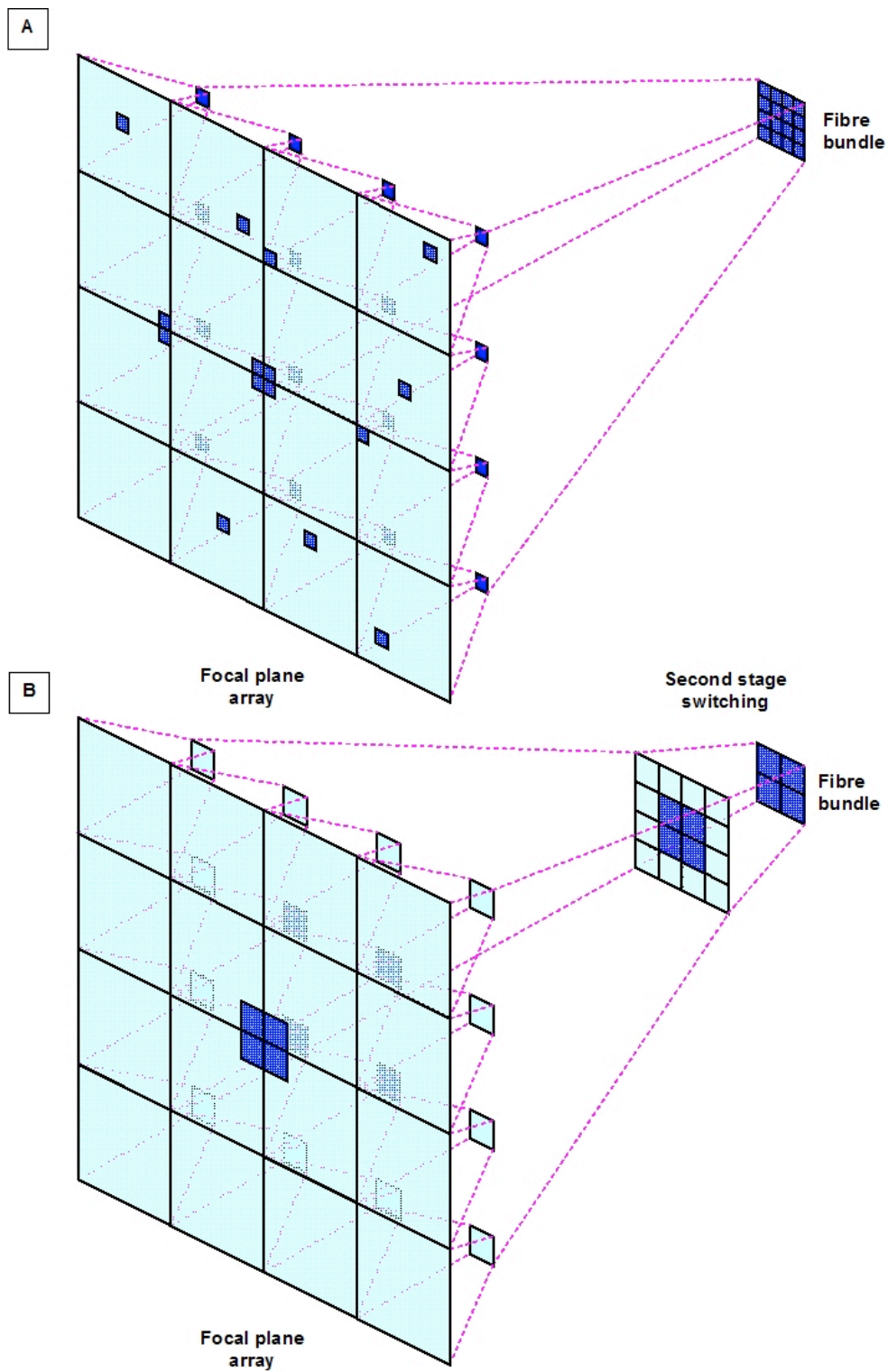

**Figure 11: Sketch illustrating the downselection using optical switch modules. In (A) the switching occurs as a single stage process. This constrains the output of the system to include equal numbers of spaxels from all patrolled sub-fields on the sky. The largest contiguous monolithic region that can be realised is at the junction where four sub-fields are adjacent. However, in (B) a second stage switching is now introduced. This decouples the output from direct interfacing with the full field. It is now possible to deselect sub-fields and choose to deploy the output fibres over a greater contiguous, monolithic region.**

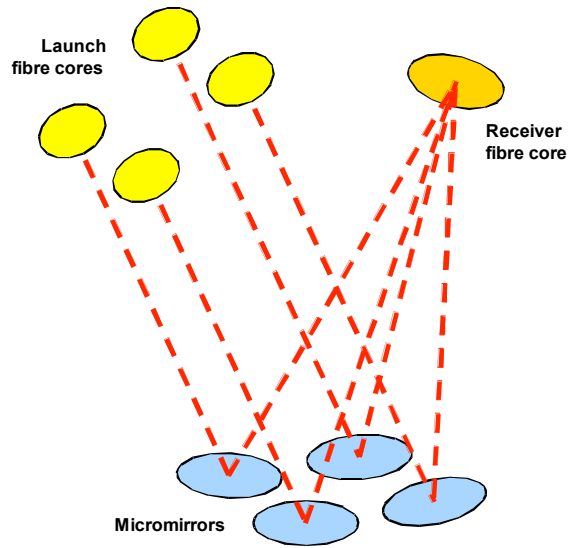

**Figure 12: Use of micromirror arrays to adapt spatial sampling by directing more than one input to a single output.**

# 5  Conclusions

We have continued the exploration of methods to provide efficient and affordable spectroscopic instrumentation on Extremely Large Telescopes with a study of techniques for highly-multiplexed spectroscopy; both the strategies and their implementation.

Cosmological studies in particular are likely to require a mixture of multiple object and integral field spectroscopy to allow the field to be addressed in an arbitrary pattern of spaxels which may be isolated or in contiguous groups: *diverse field spectroscopy.* We show an example where addressing the full field with a single large integral field unit would be inefficient compared with the diverse-field approach.

Mindful of the limitations of mechanical field-selector deployment in delivering sufficient stability, we have examined fibre-based options without large-scale mechanisms to down-select from a fibre system covering a large field of view to a smaller number of fibres which can be fed into a number of spectrographs. These may use fibre-switching networks already available commercially.

Such solutions generally require an incoherent mapping between the sky and the multiplex switcher, so that spaxels which are adjacent on the sky can be independently routed to the spectrographs as required by DFS. We examine different mapping strategies and make detailed simulations of their performance for different degrees of clustering of the regions of interest to be addressed in Paper III. This clearly demonstrates the advantage of the incoherent remapping.

In general we have demonstrated how the versatility of fibres can be exploited to serve the needs of future astronomy in a practical and efficient manner.

## Acknowledgements

We thank Miles Padgett, John Girkin and Gordon Love for their work on the bio-medical IFS applications which inspired some of this work. We also thank Anne-Marie Weijmans and Mark Swinbank for making their SAURON data available to us prior to publication and Claire Poppett for analysing mapping strategies for that field. We also thank Joss Bland-Hawthorn for suggesting improvements to the paper.